# DISCUSSION OF "2004 IMS MEDALLION LECTURE: LOCAL RADEMACHER COMPLEXITIES AND ORACLE INEQUALITIES IN RISK MINIMIZATION" BY V. KOLTCHINSKII

By A. B. Tsybakov[1]

*Université Paris 6*

The paper of Vladimir Koltchinskii has been circulating around for several years and already has become an important reference in statistical learning theory. One of the main achievements of the paper (further abbreviated as [VK]) is to propose very general techniques of proving oracle inequalities for excess risk under a control of the variance, that is, for example, under conditions (6.1) or (6.2) (often called margin or low noise conditions) or similar assumptions in terms of $L_2$-diameters $D_P(\mathcal{F}, \delta)$ and other related characteristics. These conditions lead to fast rates for the excess risk, that is, to rates that are faster than $n^{-1/2}$. The setup in [VK] is classical: methods based on empirical risk minimizers (ERM) $\hat{f}_n$ are studied under the bounded loss functions.

My comments and questions will be mainly about optimality of the excess risk bounds. This issue is not at all obvious, even in the case where the underlying class $\mathcal{F}$ is finite. We assume in what follows that either $\mathcal{F} = \{f_1, \ldots, f_M\}$, where $f_j$ are some functions on $S$, or this class is a convex hull $\mathcal{F} = \text{conv}\{f_1, \ldots, f_M\}$. Such classes $\mathcal{F}$ are used in aggregation problems where the functions $f_j$ are viewed either as "weak learners" or as some preliminary estimators constructed from a training sample which is considered as frozen in further analysis.

Let $Z_1, \ldots, Z_n$ be i.i.d. random variables taking values in a space $\mathcal{Z}$, with common distribution $P$, and denote by $\mathcal{F}_0$ the space where the $f_j$ live. Consider a loss function $Q: \mathcal{Z} \times \mathcal{F}_0 \to \mathbb{R}$ and the associated risk

$$R(f) = \mathbb{E}Q(Z, f)$$

assuming that the expectation $\mathbb{E}Q(Z, f)$ is finite for all $f \in \mathcal{F}_0$ where $Z$ has the same distribution as $Z_i$. Introduce two oracle risks: $R_{\text{MS}} = \min_{1 \le j \le M} R(f_j)$

Received March 2006.

[1] Supported by Miller Institute for Basic Research in Science, University of California, Berkeley.

corresponding to model selection-type aggregation (MS-aggregation), and $R_\text{C} = \inf_{f \in \text{conv}\{f_1,\ldots,f_M\}} R(f)$ corresponding to convex aggregation (C-aggregation). The excess risk of a statistic $\tilde{f}_n(Z_1,\ldots,Z_n)$ is defined by

$$\mathcal{E}(\tilde{f}_n) = \mathbb{E}\{R(\tilde{f}_n)\} - R_\text{OR}$$

where the oracle risk $R_\text{OR}$ equals either $R_\text{MS}$ or $R_\text{C}$. A natural question about optimality is the following: how to find an estimator $\tilde{f}_n$ for which the excess risk is as small as possible? Of course, this question cannot be answered simultaneously for all distributions $P$. However, optimality can be treated in a minimax sense: introduce a class $\mathcal{P}$ of distributions and call $\varepsilon_{n,M}$ *optimal rate of aggregation* if

$$\inf_{T_n} \sup_{\mathcal{G}_M} \mathcal{E}(T_n) \asymp \varepsilon_{n,M}$$

where $\mathcal{G}_M = \{(P, f_1, \ldots, f_M) : P \in \mathcal{P},\ f_j \in \mathcal{F}_0\}$ and the infimum is taken over all estimators $T_n$. An estimator $\tilde{f}_n$ is declared to be optimal if it achieves

$$\sup_{\mathcal{G}_M} \mathcal{E}(\tilde{f}_n) \leq C \varepsilon_{n,M}$$

for some constant $C$ independent of $n$ and $M$. Optimal rates of aggregation are known for several important special cases [6]: for instance, if $\mathcal{F}_0$ is the class of all functions bounded in absolute value by a given constant, in Gaussian (or bounded) regression model with squared loss the optimal rates are

$$(1) \quad \varepsilon_{n,M} \asymp \begin{cases} M/n, & \text{for C-aggregation if } M \leq \sqrt{n}, \\ \sqrt{\dfrac{1}{n} \log\left(\dfrac{M}{\sqrt{n}} + 1\right)}, & \text{for C-aggregation if } M > \sqrt{n}, \\ (\log M)/n, & \text{for MS-aggregation,} \end{cases}$$

and optimal procedures $\tilde{f}_n$ attaining these rates are available [6].

The paper [VK] suggests very general bounds on probabilities of deviations of $R(\hat{f}_n) - R_\text{OR}$ where $\hat{f}_n$ is an empirical risk minimizer. Clearly, these bounds can be applied to evaluate the expected risk $\mathcal{E}(\hat{f}_n)$ and to check whether $\hat{f}_n$ attains optimality (at least for the bounded regression model). I think that this should be the case for C-aggregation, probably under some more assumptions on $n$ and $M$, but in general not for MS-aggregation. Furthermore, presumably no selector, that is, no procedure that chooses only one of the $M$ functions as estimator, can achieve the MS-rate given in (1) under strictly convex loss. On the other hand, MS-optimality can be achieved by estimators $\tilde{f}_n$ that are convex mixtures of $f_1, \ldots, f_M$ with data-dependent coefficients. A simple aggregation method of this kind called *mirror averaging* [3, 4] is defined as follows.

Let $\Lambda^M = \{\theta = (\theta^{(1)}, \ldots, \theta^{(M)}) : \theta^{(j)} \geq 0, \sum_{j=1}^M \theta^{(j)} = 1\}$ be the unit simplex in $\mathbb{R}^M$, and let $G : \mathbb{R}^M \to \Lambda^M$ be a function satisfying certain assumptions [3]. A possible choice of $G$ is

$$G(z) = \left(\frac{\exp(-z^{(1)})}{\sum_{j=1}^M \exp(-z^{(j)})}, \ldots, \frac{\exp(-z^{(M)})}{\sum_{j=1}^M \exp(-z^{(j)})}\right),$$

$z = (z^{(1)}, \ldots, z^{(M)})$. This particular function $G$ (corresponding to the Gibbs distribution) will be considered in what follows. To any $z \in \mathbb{R}^M$ we associate its "mirror image" in the simplex $\Lambda^M$, that is, a probability vector $G(z/\beta)$ where $\beta > 0$ is a tuning parameter.

For any $\theta = (\theta^{(1)}, \ldots, \theta^{(M)}) \in \Lambda^M$ set $\mathsf{f}_\theta = \sum_{j=1}^M \theta^{(j)} f_j$ and assume that $Q(Z, \mathsf{f}_\theta)$ is differentiable w.r.t. $\theta$ with gradient $\nabla_\theta Q(Z, \mathsf{f}_\theta)$. Given two sequences of positive numbers $\beta_i$ and $\gamma_i$, the mirror averaging (MA) algorithm is defined as follows:

- $i = 0$: initialize values $\zeta_0 \in \mathbb{R}^M$, $\bar{\theta}_0 \in \Lambda^M$, $\tilde{\theta}_0 = 0$,
- for $i = 1, \ldots, n$, iterate:

$$\zeta_i = \zeta_{i-1} + \gamma_i \nabla_\theta Q(Z_i, \mathsf{f}_{\bar{\theta}_{i-1}}) \quad \text{(GRADIENT DESCENT)}$$

$$\bar{\theta}_i = G(\zeta_i/\beta_i) \quad \text{(MIRRORING)}$$

$$\tilde{\theta}_i = \frac{\sum_{t=1}^i \gamma_t \bar{\theta}_{t-1}}{\sum_{t=1}^i \gamma_t} \quad \text{(AVERAGING)}$$

- output $\tilde{\theta}_n$ and set $\tilde{f}_n = \mathsf{f}_{\tilde{\theta}_n}$.

Remark that the vector of weights $\tilde{\theta}_n$ belongs to the simplex $\Lambda^M$, so that $\tilde{f}_n$ is a convex mixture of initial functions (estimators) $f_j$ with data-dependent weights. The following theorem proved in [3] shows that the MA estimator satisfies a sharp oracle inequality.

THEOREM 1 (Convex aggregation). *Let $\theta \mapsto Q(Z, \mathsf{f}_\theta)$ be convex on $\Lambda^M$ for all $Z \in \mathcal{Z}$ and*

(2) $$\sup_{\theta \in \Lambda^M} \mathbb{E}\|\nabla_\theta Q(Z, \mathsf{f}_\theta)\|_\infty^2 \leq Q^\star$$

*where $\|\cdot\|_\infty$ is the sup-norm in $\mathbb{R}^M$. Then the mirror averaging algorithm with appropriate $\beta_i$ and $\gamma_i$ outputs $\tilde{f}_n$ that satisfies*

(3) $$\mathbb{E}\{R(\tilde{f}_n)\} - R_C \leq 2\sqrt{Q^\star}\sqrt{\frac{\log M}{n}}$$

*for all $n \geq 1$, $M \geq 2$.*

If $Q^\star$ does not depend on $M$, the rate of convergence on the right-hand side of (3) is optimal for $M$ comfortably larger than $\sqrt{n}$ [cf. (1)]. Although this is not explicitly stated in [VK], it seems that a similar result can be obtained for the ERM $\hat{f}_n$ if $Q$ is strictly convex in $\theta$ (condition (7.5) of [VK]) using the techniques of Sections 7 and 8. In particular, the second statement of Theorem 13 covers the case of squared loss $Q$. It would be interesting to compare these developments to (3) and to check whether, for a general class of convex functions, the ERM or MA estimators achieve optimality in the zone $M \leq \sqrt{n}$ where the bound of Theorem 1 is suboptimal. Note that, in a difference with [VK], Theorem 1 is not restricted to bounded loss functions or to loss functions with bounded gradient. Moment conditions on the components of the gradient suffice, but lead to coarser bounds where $Q^\star$ grows with $M$.

A particular instance of the MA algorithm can be used to mimic the MS-oracle with sharp bounds on the excess risk. It is called the *linearized mirror averaging* (LMA) algorithm and is defined in the same way as MA, with the only difference that the gradient descent step is modified as follows [4]:

$$\zeta_i = \zeta_{i-1} + u_i \qquad \text{where } u_i = (Q(Z_i, f_1), \ldots, Q(Z_i, f_M))^\top.$$

Thus, LMA is a special case of mirror averaging associated to the "surrogate" linear risk $Q^L(Z, \theta) = \theta^\top u(Z)$ where $u(Z) = (Q(Z, f_1), \ldots, Q(Z, f_M))^\top$. Two special cases of LMA, for the regression with squared loss and for density estimation with Kullback loss, have been studied earlier (cf. [2] and the references therein).

To state a general excess risk bound for LMA, introduce the random variable $\omega$ taking values $1, \ldots, M$ with the distribution $\mathbf{P}$ defined conditionally on $(Z_1, \ldots, Z_n)$ by $\mathbf{P}(\omega = j) = \tilde{\theta}_n^{(j)}$ where $\tilde{\theta}_n^{(j)}$ is the $j$th component of $\tilde{\theta}_n$. The expectation corresponding to $\mathbf{P}$ is denoted by $\mathbf{E}$. The following bound is proved in [4].

THEOREM 2 (MS). *Let $\tilde{\theta}_n$ be the output of LMA algorithm with $\beta_i \equiv \beta > 0$, $\gamma_i \equiv 1$, and let the loss function $Q$ be such that*

$$(4) \qquad \mathbb{E}\log\left(\mathbf{E}\exp\left[\frac{Q(Z, \mathbf{E}[\omega]) - Q(Z, \omega)}{\beta}\right]\right) \leq 0,$$

*where $\mathbb{E}$ denotes the expectation w.r.t. the joint distribution of $n + 1$ i.i.d. random variables $(Z_1, \ldots, Z_n, Z)$. Then*

$$(5) \qquad \mathbb{E}\{R(\tilde{f}_n)\} - R_{\text{MS}} \leq \frac{\beta \log M}{n + 1}.$$

Condition (4) is satisfied for loss functions $Q$ that are "in the average" (or approximately, up to a set in $\mathcal{Z}$ of small measure) strongly convex in $\theta$;

several sufficient conditions for (4) can be found in [4]. The most simple of them is concavity of the mapping

$$\theta \mapsto \mathbb{E} \exp\left(\frac{Q(Z,\theta') - Q(Z,\theta)}{\beta}\right) \quad (6)$$

on the simplex $\Lambda^M$ for any fixed $\theta' \in \Lambda^M$. We will say that a loss function is *nice* if it satisfies (4). In contrast to the convex aggregation bound considered in Theorem 1, inequality of the form (5) for the excess risk is not obtained for empirical risk minimizers, and I conjecture that it is not true for them without additional strong restrictions on $P$ such as $R_{\rm MS} = 0$ or the margin assumption with parameter $\kappa = 1$.

Note that on the right-hand side of (5) we have the optimal rate of MS-aggregation, which proves that the LMA procedure is rate optimal for *nice* loss functions. To see how sharp the bound (5) is, consider the classification model with convex loss. Let $Z = (X,Y)$ where $X \in \mathbb{R}^d$ is a random predictor and $Y \in \{-1,1\}$ is a random label. Assume that we have $M$ classifiers $f_j : \mathbb{R}^d \to [-1,1]$, $j = 1, \ldots, M$. Consider the loss function $Q(Z,f) = \varphi(-Yf(X))$ where $\varphi : \mathbb{R} \to \mathbb{R}_+$ is a convex twice differentiable function. The associated risk is the $\varphi$-risk of classification:

$$R(f) = \mathbb{E}\varphi(-Yf(X)). \quad (7)$$

Then the mapping (6) is concave if $(\varphi'(x))^2 \le \beta \varphi''(x)$, $\forall\, |x| \le 1$. This implies, for example, that inequality (5) holds for $R$ of the form (7) with rather sharp constants: $\beta = e$ if $\varphi(x) = e^x$ (exponential boosting) and $\beta = e \log 2$ if $\varphi(x) = \log_2(1+e^x)$ (logit boosting). It would be interesting to study whether these constants can be improved by any estimation method.

Finally, let me mention some other open problems related to optimality of excess risk bounds.

(I) Theorem 1 holds for convex loss and Theorem 2 for nice (essentially, strongly convex) loss. What are optimal excess risk bounds for other loss functions?
(II) What are optimal excess risk bounds over restricted classes of underlying distributions, for example, under control of the variance (such as the low noise, or margin, assumption)?
(III) Theorems 1 and 2 deal with two simple classes $\mathcal{F}$: finite classes and their convex hulls. How to treat general classes $\mathcal{F}$? What are optimal rates of aggregation [analog of (1)] for general $\mathcal{F}$?

Theorem 12 in [VK] gives an insight into (III). It considers the class $\mathcal{F}$ which is a convex hull of a $V$-dimensional set. Instead of the number of functions $M$ (in our case), the key parameter in Theorem 12 is the metric dimension or the VC-dimension $V$. Theorem 12 gives only an upper bound. How optimal is it?

Some first results for the problems (I) and (II) have been recently obtained by Lecué [5]. He considers classification setup as stated above, with $\varphi$ being either the hinge loss or the indicator loss, for the class of distributions $P$ satisfying the margin assumption (cf. (6.2) in [VK]) with exponent $\kappa \geq 1$, and he suggests aggregate classifiers $f_n^\star$ such that

$$(8) \quad \mathbb{E}\{R(f_n^\star)\} \leq R_{\mathrm{MS}} + C\left[\sqrt{\frac{(R_{\mathrm{MS}} - R_\star)^{1/\kappa} \log M}{n}} + \left(\frac{\log M}{n}\right)^{\frac{\kappa}{2\kappa-1}}\right]$$

where $R$ is either the hinge risk or the probability of misclassification, $R_{\mathrm{MS}}$ is the corresponding MS-oracle risk, $R_\star$ is the risk of Bayes classifier and $C > 0$ is a constant. Furthermore, [5] proves a minimax lower bound showing that the expression in square brackets in (8) plays the role of optimal rate, analogous to $\varepsilon_{n,M}$. It is interesting that, in contrast to the bounds of Theorems 1 and 2, here the optimal rate depends not only on $n$ and $M$, but also on the difference between the oracle risk and the risk of Bayes classifier. Note also that, for the hinge risk, C-aggregation is identical to MS-aggregation since for classifiers taking values in $[-1, 1]$ we have $R_{\mathrm{MS}} = R_{\mathrm{C}}$. The optimal rate of MS-aggregation cannot be as fast as for *nice* loss functions [cf. (1)], except for the most favorable case where $\kappa = 1$. These remarks show that what we should expect to get in (I) and (II) is quite different from the previously obtained results.

My last question falls somewhat apart from the above discussion. Consider again the classification problem under the margin condition with exponent $\kappa > 1$, and assume that the regression function $\eta$ belongs to a class of functions with the $L_\infty$ log-covering number of the order $\varepsilon^{-\rho}$, $\rho > 0$ (such as a Hölder or Sobolev class). The last assumption is natural when plug-in classifiers, in particular, the SVM or boosting-type ones are studied. The optimal rate of convergence of the excess Bayes risk under these assumptions is a (potentially fast) rate of the order $\psi_n \triangleq n^{-\frac{\kappa}{2\kappa-1+\rho(\kappa-1)}}$ [1]. The ERM classifiers attaining this rate suggested in [1] are based on $L_\infty$-covering of the set of regression functions $\eta$, while the argument in [VK] uses $L_2$-covering of the set of indicators $f(\cdot) = I\{\eta(\cdot) \geq 1/2\}$, which apparently leads to slower rates. Can this argument be extended to prove that the ERM attains the optimal rate $\psi_n$?

Laboratoire de Probabilités
et Modèles aléatoires
UMR 7599, Université Paris 6
case 188, 4, pl. Jussieu
F-75252 Paris Cedex 5
France
E-mail: tsybakov@ccr.jussieu.fr